\documentstyle[12pt]{article}

\begin{document}
\noindent
\begin{large}
{\bf Comment on semi-classical approaches to giant magnetoresistance in
 magnetic multilayers}
 \end{large}

 The semi-classical approaches to giant magnetoresistance
 observed in magnetic
multilayers originated from
 Camley and Barn\'{a}s \cite{camley}\cite{c2} achieved success.  I point out
 here some issues worthy noticing.

The spin state of one electron can be
expressed as a linear superposition of a complete orthonormal set of
eigenstates,
the up and down states
along a spin
quantization axis,
$|\psi>\,=\,c_{\uparrow}|\uparrow>+c_{\downarrow}|\downarrow>$ with
 $|c_{\uparrow}|^{2}\,+\,|c_{\downarrow}|^{2}\,=\,1$ . For
{\em stationary state}, $|c_{\uparrow}|^{2}$ and $|c_{\downarrow}|^{2}$
are independent of time. Equivalently, the system of many electrons can
be described by that there are $|c_{\uparrow}|^{2}$ in up state and
$|c_{\downarrow}|^{2}$ in down state. The distribution function may be
divided consequently into two parts by multiplying $|c_{\uparrow}|^{2}$
and $|c_{\downarrow}|^{2}$. This was actually what Refs. \cite{camley}
and \cite{c2}
did.

Alternatively $|\psi>$ can also be superposed by another
complete orthonormal
set,
$|\psi>\,=\,c_{\uparrow}'|\uparrow'>+c_{\downarrow}'|\downarrow'>$.
If the solid angle between the latter and the former  quantization axes
is $(\theta,\phi)$, we have
\begin{equation}
|\uparrow'>\,=\,\cos(\theta/2)\exp(-i\phi/2)|\uparrow>+
   \sin(\theta/2)\exp(i\phi/2)|\downarrow>,
\end{equation}
\begin{equation}
|\downarrow'>\,=\,\sin(\theta/2)\exp(-i\phi/2)|\uparrow>-
   \cos(\theta/2)\exp(i\phi/2)|\downarrow>.
\end{equation}
The expansion coefficients  are related by
\begin{equation}
c_{\uparrow}\,=\,c_{\uparrow}'\cos(\theta/2)\exp(-i\phi/2)+c_{\downarrow}'
\sin(\theta/2)\exp(-i\phi/2),
\end{equation}
\begin{equation}
c_{\downarrow}\,=\,c_{\uparrow}'\sin(\theta/2)\exp(i\phi/2)-c_{\downarrow}'
\cos(\theta/2)\exp(i\phi/2).
\end{equation}
Therefore
\begin{equation}
|c_{\uparrow}|^{2}\,=\,\cos^{2}(\theta/2)|c_{\uparrow}'|^{2}
+\sin^{2}(\theta/2)|c_{\downarrow}'|^{2}
+Re[(c_{\uparrow}')^{*}c_{\downarrow}']\sin\theta,  \label{eq:e1}
\end{equation}
\begin{equation}
|c_{\downarrow}|^{2}\,=\,\sin^{2}(\theta/2)|c_{\uparrow}'|^{2}
+\cos^{2}(\theta/2)|c_{\downarrow}'|^{2}
-Re[(c_{\uparrow}')^{*}c_{\downarrow}']\sin\theta,     \label{eq:e2}
\end{equation}

Camley and Barn\'{a}s introduced a fictitious plane in the nonmagnetic layer
where the quantization axis changes from  the magnetization   of one
ferromagnetic layer to another, this is nothing but that decompose there the
spin state of one electron in two different complete orthonormal sets.
Comparing with Eqs. (\ref{eq:e1}) and (\ref{eq:e2}) above, we see
that the last terms, that of interference  were lost in Refs. \cite{camley}
and \cite{c2}. The Eqs. (9) to (12) in Ref. \cite{camley} or
Eqs. (6) to (9) in Ref.\cite{c2} are wrong for a general $\theta$.
Fortunately and accidently, when and only when $\theta\,=\,0$ or $\pi$,
the interference terms disappear hence  Eqs. (\ref{eq:e1}) and
(\ref{eq:e2}) simplify to
$|c_{\uparrow}|^{2}$ = $|c_{\uparrow}'|^{2}$ and $|c_{\downarrow}|^{2}$ =
$|c_{\downarrow}|^{2}$ if $\theta\,=\,0$ while $|c_{\uparrow}|^{2}$ =
$|c_{\downarrow}'|^{2}$ and $|c_{\downarrow}|^{2}$ = $|c_{\uparrow}'|^{2}$
if $\theta\,=\,\pi$. This is why one can only consider
the correspondences between majority and minority
spins and up and down spins \cite{hood}.

The configuration with a general angle between neighboring magnetization
is possible, as seen from the gradual change of magnetoresistance when
the external magnetic field was along the hard axis in the experiment
by Binasch {\it et al.} \cite{binasch}.
This is beyond the capability of semiclassical
approaches dealing only distrubution function since the interference terms
do not vanish generally.
This can be addressed by quantum approach with the consideration here taken
into account \cite{shi}.

This work is supported by China Natural Science Foundation
under Grant No. 19374013.
Prof. Ruibao Tao, Lei Zhou and Baoxing Chen  are thanked for valuable
discussions.

\newpage
{\noindent}
Yu Shi

Department of Physics

Fudan University

Shanghai 200433, P. R. China
\vspace{1cm}


\end{document}